\documentclass[twocolumn,showpacs,preprintnumbers,amsmath,amssymb,floatfix,prd]{revtex4}
\usepackage{epsfig}
\usepackage{dcolumn}
\begin{document}
\title{The role of heavy quarks in light hadron fragmentation.}
\author{Manuel Epele}\email{manuepele@fisica.unlp.edu.ar}
\affiliation{ Instituto de F\'{\i}sica La Plata, UNLP, CONICET 
Departamento de F\'{\i}sica,  Facultad de Ciencias Exactas, Universidad de
La Plata, C.C. 69, La Plata, Argentina}
\author{Carlos Garc\'{\i}a Canal}\email{garcia@fisica.unlp.edu.ar }
\affiliation{ Instituto de F\'{\i}sica La Plata, UNLP, CONICET 
Departamento de F\'{\i}sica,  Facultad de Ciencias Exactas, Universidad de
La Plata, C.C. 69, La Plata, Argentina}
\author{R. Sassot}\email{sassot@df.uba.ar} 
\affiliation{Departamento de F\'{\i}sica and IFIBA,  Facultad de Ciencias Exactas y
 Naturales, Universidad de Buenos Aires, Ciudad Universitaria, Pabell\'on\ 1 (1428) 
Buenos Aires,  Argentina}
\begin{abstract}
We investigate the role of heavy quarks 
in the production of light flavored hadrons and in the determination of the corresponding 
non perturbative hadronization probabilities. 
We define a general mass variable flavor number scheme for fragmentation functions
that accounts for heavy quark mass effects,
and perform a global QCD analysis to an up-to-date data set including very precise Belle 
and BaBar results.  We show that the mass dependent picture provides a much more accurate 
and consistent description of data.
\end{abstract}


\maketitle
%
%
{\it Introduction.---} The effects of heavy quark masses in hard processes and the 
appropriate definition of parton probability densities for such species of quarks 
have been very actively studied in recent years \cite{Ball:2015dpa}.
When a heavy quark 
participates in a hard process, and the characteristic energy of the process under consideration is not far from the heavy quark 
mass scale, the most natural choice is to treat these particles as massive throughout the 
calculation, rather than appeal to the more conventional massless parton approximation. 
However, when the scale of the process exceeds by large the mass scale of the heavy quarks, 
mass corrections not only become negligible, but the all-order resummations implicit in 
massless parton approaches become crucial. This situation clearly represents a challenge 
for a precise and consistent comparison of processes occurring at very different energy 
scales, as it necessarily happens in a global QCD analysis designed to extract non-perturbative 
parton distribution functions (PDFs) \cite{Rojo:2015acz} or fragmentation functions (FFs) 
\cite{deFlorian:2007aj,ref:other-ffs} from the data. 
To overcome this problem, in most modern global QCD analysis of data, the so called general 
mass variable flavor number (GMVFN) schemes \cite{Aivazis:1993pi,Thorne:2000zd,Thorne:2006qt,Guzzi:2011ew} for parton densities are introduced, as they 
allow to retain the advantages of massive schemes near the mass thresholds and those of the 
massless approach at high energies, smoothly interpolating between both regimes. 

In fact, different GMVFN approaches have been applied to the analysis of fragmentation 
probabilities of heavy quarks into heavy flavored hadrons \cite{Cacciari:2005ry,Kniehl:2008zza,Kneesch:2007ey}, but until now little 
attention has been paid in this respect to light hadron fragmentation processes. Since the 
charm and bottom content in the proton is rather limited, the production of light mesons 
via heavy quarks of course is strongly suppressed relative to that via light quarks both 
in proton-proton collisions and in semi-inclusive deep inelastic scattering (SIDIS). 
Heavy quark corrections are thus expected to be negligible for these processes. But that 
is not the case for single inclusive electron-positron annihilation (SIA) into light 
mesons, where the charm and  bottom contribution to the cross section is estimated to be 
comparable in size to that of the light flavors \cite{deFlorian:2014xna}. Whereas charm 
and bottom mass corrections may still be negligible in SIA 
experiments tuned at the mass of the Z boson, they are certainly relevant at the energy 
scales of the more recent BaBar and Belle experiments \cite{ref:babardata,ref:belledata}, 
which are just above the bottom mass threshold.   

In this paper we compute the single inclusive electron-positron annihilation cross section 
with non zero masses for the charm and bottom quarks at first order in the strong coupling 
constant $\alpha_s$, and we estimate the effects of retaining these mass corrections in
pion production. 
Since we find the mass dependence to be large but we want to keep the advantages of the 
NLO massless approximation at high energies, we define a general mass variable flavor 
number scheme for fragmentation functions in the lines of the FONLL scheme \cite{Forte:2010ta,Ball:2015tna}, commonly 
used for PDFs. We implement numerically this scheme in Mellin 
moment space for the fast computation of the cross sections as required by QCD global 
analyses, and we perform them including recent Belle and BaBar data. 
The mass dependent picture introduced by 
the GMVFN scheme is found 
to be relevant in the extraction of fragmentation functions both in terms of 
the quality of the fit to data and in the reduction of the 
normalization shifts applied to data that are customarily included in global fits.
The shape of the charm into pion fragmentation function, that contributes significantly to 
the cross section at the energies of the Belle and BaBar experiments, is noticeably modified 
relative to the results obtained within a massless scheme. The bottom fragmentation, 
constrained mainly by higher energy data, remains similar to the one found in the massless 
approximation. 


{\it Factorization Schemes.---} Most analyses of quark fragmentation into light flavored 
hadrons \cite{deFlorian:2007aj,ref:other-ffs} rely on the massless perturbative QCD approximation, supplemented with 
heavy quark mass thresholds, where the corresponding heavy quarks become active,
contribute to cross sections and enter the scale or evolution equations \cite{ref:dglap}.  
This zero mass variable flavor number (ZMVFN) scheme 
is the simplest framework to compute the SIA cross section \cite{ref:aempi}:
\begin{equation}
  \frac{ d\sigma}{dz}^{\text{ZMVFN}} = \sum _{i = q,g,h}\hat{\sigma}_i^{\text{ZM}}(z,Q) 
    \otimes D_i^{\text{ZM}}(z,Q)  \label{masslesscs}
\end{equation}
where $\hat{\sigma}_i^{\text{ZM}}$ is the massless partonic SIA cross 
section into a parton of flavor $i$, and $D_i^{\text{ZM}}$ are the corresponding FFs,
for which we omit in what follows the dependence on the scaled hadron energy fraction $z$.
$\otimes$ represent the appropriate convolution over $z$, and $Q$ is the center of mass energy. 
$D_i^{\text{ZM}}$ evolves in the scale $Q$ through massless QCD 
evolution equations for any parton flavor $i$, that include light ($q$) and heavy ($h$) 
quarks and antiquarks and gluons ($g$). Eq.(1) gives a remarkably good approximation
at NLO and NNLO well above mass threshold $m_{h}$ \cite{deFlorian:2014xna,Anderle:2015lqa}, 
i.e. $Q \gg m_{h}$, but fails to account for mass effects. Alternatively, in a massive scheme (M), heavy quark masses are kept
at the partonic cross section level,
\begin{eqnarray}
  \frac{ d\sigma}{dz}^{\text{M}} \hspace{-7pt} = \hspace{-3pt} \sum _{i = q,g} \hspace{-1pt} \hat{\sigma}_i^{\text{M}}(Q,m_{h}) \otimes D_i^{\text{M}}(Q) + \hat{\sigma}_h^{\text{M}}(Q,m_{h}) \otimes D_h^{\text{M}}
    \label{massivecs}
\end{eqnarray}
the light flavored FFs $D_{q,g}^{\text{M}}(Q)$ still evolve through 
massless QCD evolution equations in NLO, although heavy quark loops contribute above 
$\mathcal{O}(\alpha_S^2)$, and heavy quark FFs $D_{h}^{\text{M}}$ decouple from the QCD 
evolution \cite{Collins:1998rz}. 
The massive scheme gives a good description near the mass thresholds, but fail to converge
to the massless limit at high energies because of potentially large logarithmic contributions 
($\alpha_{S}^k\log^k(m_h/Q)$) present in the partonic cross sections $\hat{\sigma}_i^{\text{M}}$ 
that spoil the convergence of the perturbation expansion. These logarithmic contributions are
effectively resummed in the renormalization group improved ZM approximation. In fact, 
it has been shown \cite{Buza:1996wv} that in the massless limit, i.e. $Q \gg m_h$, 
\begin{equation}
    \hat{\sigma}_i^{\text{M}}(Q,m_h) \xrightarrow[m_h\rightarrow0]{} \sum _{j = q,g,h} \hat{\sigma}_j^{\text{ZM}}(Q) \otimes \mathcal{A}_{ji}(Q/m_h)
    \label{massfactorization}
\end{equation}
where all logarithmic contributions can be factorized in an operator matrix 
$\mathcal{A}_{ij}$ that is independent of the hard process under consideration 
\cite{Kniehl:2004fy,Kniehl:2008zza,Kneesch:2007ey}.

\begin{figure}[hbt!]
\vspace*{-0.4cm}
\hspace*{-6mm}
\epsfig{figure=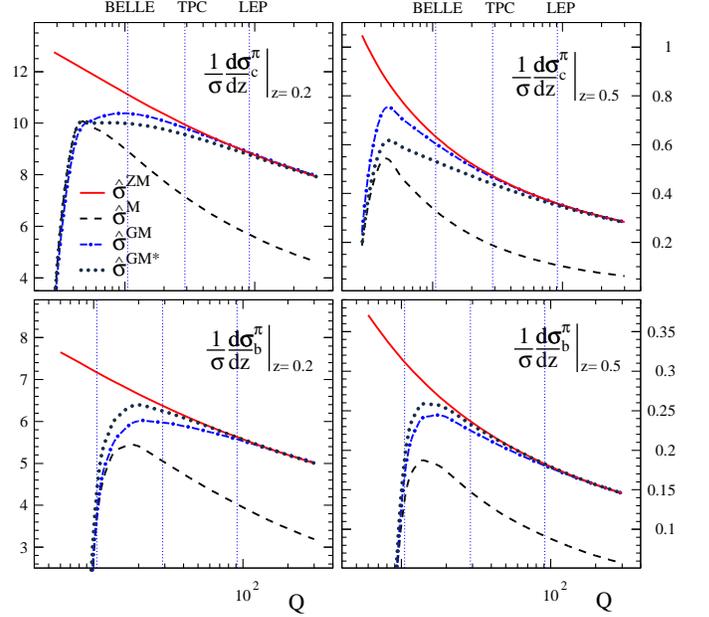,width=0.54\textwidth}
\vspace*{-0.5cm}
\caption{Charm and bottom contribution to the SIA cross section computed with different 
approximations for the partonic cross sections and the same set of FFs.}
\label{fig:cb_evo}
\end{figure}

The advantages of both the zero mass and the massive schemes can be exploited in a general 
mass (GM) scheme, which defines the corresponding fragmentation probabilities through
\begin{equation}
  \frac{ d\sigma}{dz}^{\text{GMVFN}} = \sum _{i = q,g,h} \hat{\sigma}^{\text{GM}}_j(Q,m_h) \otimes D_j^{\text{GM}}(Q)
    \label{matchingcs}
\end{equation}
where a subtracted massive partonic cross section
\begin{equation}
    \hat{\sigma}_j^{\text{GM}}(Q,m_h) = \sum _{i = q,g,h} \hat{\sigma}_i^{\text{M}} (Q,m_h) \otimes \mathcal{A}_{ij}^{-1}(Q/m_h)
    \label{matchingpartoncs}
\end{equation}  
guarantees the correct massless behavior at high energies. The fragmentation functions
$D_j^{\text{GM}}(Q)$ obey standard evolution equations as in the ZM scheme and the 
continuity across the thresholds can be ensured by imposing the following matching 
condition
\begin{equation}
    D_j^{\text{GM}}(m_h) = \sum _{i = q,g,h} \mathcal{A}_{ji}(1) \otimes {D}_i^{\text{M}}(m_h)
    \label{matchingcondition}
\end{equation}
analogously as in the FONLL scheme for PDFs \cite{Forte:2010ta,Ball:2015tna}.

In order to illustrate the effects arising from the use of the different approximations 
for the partonic cross sections, in Fig.\ref{fig:cb_evo} we show the charm and bottom 
contribution to the SIA pion cross section (upper and lower panels, respectively)
as a function of the center of mass energy $Q$ at two reference values of $z$. 
For the comparison, the same set of NLO FFs that will be described in detail in the next 
section, is convoluted with the NLO massless (ZM) partonic cross section (red solid lines), 
the full $\mathcal{O}(\alpha_S)$ massive (M) result (black dashed lines), and the subtracted 
(GM) approximation (blue dot-dashed lines). Even though the scheme defined by 
Eqs.(\ref{matchingcs}-\ref{matchingcondition}) interpolates between the massive behavior 
near thresholds and the desired zero mass limit at $Q \gg m_h$, this solution is clearly 
not unique. The same limits can be satisfied in alternative factorization schemes where 
the convergence to the massless limit happens at a lower or higher energy scale. 
For instance, substituting
\begin{equation}
  \hat{\sigma}^{\text{GM}}_j \longrightarrow \hat{\sigma}^{\text{GM*}}_j =  (1 -f(Q))\hspace{5pt}\hat{\sigma}_j^{\text{M}} + f(Q)\hspace{5pt}\hat{\sigma}^{\text{GM}}_j
    \label{massivemix}
\end{equation}
in eq.(\ref{matchingcs}), with an $f(Q)$  that vanishes in the threshold $f(m_h) = 0$,
and saturates to 1 for $Q \gg m_h$ such as $f(Q)=1-2m_h/Q$, would delay the onset of 
the massless-like behaviour, as shown for the charm contribution by the grey dotted 
lines in the upper panels of Fig.\ref{fig:cb_evo}, while
\begin{equation}
 \hat{\sigma}^{\text{GM}}_j \longrightarrow \hat{\sigma}^{\text{GM*}}_j =  (1 -f(Q))\hspace{5pt}\hat{\sigma}^{\text{GM}}_j + f(Q)\hspace{5pt}\hat{\sigma}_j^{\text{ZM}}
    \label{masslessmix}
\end{equation} 
would suppress mass effects, as shown for the bottom contribution in the lower panels.
In any case, it is clear that while the GM schemes introduce negligible corrections 
at the energy scale of the LEP experiments, mass effects are significant at the 
scales of Belle and BaBar. The remarkably precise measurements of the SIA cross section
as a function of $z$ performed by these experiments allows to check if a particular 
scheme is favored by the data. 
 
{\it GMVFN scheme global analysis for FFs.---} 
In this section, we discuss the actual relevance of heavy quark mass corrections 
in charged pion production, implementing different factorization schemes in a NLO QCD 
global analysis for the extraction of FFs, performed along the lines of that of ref. 
\cite{deFlorian:2014xna}. The method for the global analysis has been described in 
detail in \cite{deFlorian:2007aj,deFlorian:2014xna}. 
It is based on an efficient Mellin moment technique that allows one to tabulate and store 
the computationally most demanding parts of the NLO calculation of SIA, SIDIS and 
proton-proton hadroproduction cross sections prior to the actual analysis. In this way, 
the evaluation of the relevant cross sections becomes so fast that can be easily performed  
inside a standard $\chi^2$ minimization. At variance with \cite{deFlorian:2014xna}, where 
SIA cross sections were evaluated only in the ZM approximation, here they are extended to 
the GM framework what requires to define different contours in the complex moment space
to perform the Mellin inversion. 
Hadroproduction and SIDIS cross sections are still computed in the ZM framework, to asses 
in this first step the impact introduced by SIA corrections. Nevertheless, the heavy quark
contributions to these processes are negligibly small. Also a minor difference with 
\cite{deFlorian:2014xna} is that we remove from the data sets included in the analysis 
TASSO and OPAL light flavor tagged data, that have comparatively large errors.   
\begin{table}[t!]
\caption{\label{tab:exppiontab} Individual $\chi^2$ values and normalization shifts 
$N_i$ for the data sets included in global analyses where the ZMVFN and GMVFN schemes 
have been implemented.}  
\begin{ruledtabular}
\begin{tabular}{lcccrcr}
%
%
experiment& data  & \# data     &   \multicolumn{2}{c}{ZMVFN}         &\multicolumn{2}{c}{GMVFN}         \\
          & type  & in fit       &   $N_i$   & $\chi^2$     &  $N_i$   & $\chi^2$   \\\hline
{\sc Aleph} \cite{ref:alephdata}    & incl.\  &  22 & 0.968 & 21.6 & 0.994 & 23.3\\
{\sc BaBar} \cite{ref:babardata}     & incl.\  & 39  & 1.019 & 76.7 & 1.002 & 58.2\\ 
{\sc Belle} \cite{ref:belledata}     & incl.\  & 78  & 1.044 & 19.5 & 1.019 & 11.0\\  
{\sc Delphi} \cite{ref:delphidata}  & incl.\  & 17 & 0.978 & 6.7  & 1.003 & 9.3\\
                     & $uds$ tag              & 17 & 0.978 & 20.8 & 1.003 & 9.5\\
                     & $b$ tag                & 17 & 0.978 & 10.5 & 1.003 & 7.8\\
{\sc Opal} \cite{ref:opaldata}  & incl.\ &  21 & 0.946 & 27.9 & 0.970 & 15.9\\ 
{\sc Sld} \cite{ref:slddata}  & incl.\  &  28 & 0.938 & 28.0 & 0.963 & 9.5\\
          & $uds$ tag                   &  17 & 0.938 & 21.3 & 0.963 & 11.3\\
          & $c$ tag                     &  17 & 0.938 & 34.0 & 0.963 & 19.8\\
          & $b$ tag                     &  17 & 0.938 & 11.1 & 0.963 & 9.9\\
{\sc Tpc} \cite{ref:tpcdata}  & incl.\         & 17 & 0.997 & 31.7 & 1.006  & 27.9\\
                              & $uds$ tag      &  9 & 0.997 & 2.0  & 1.006  & 2.0\\
                              & $c$ tag        &  9 & 0.997 & 5.9  & 1.006  & 4.3\\
                              & $b$ tag        &  9 & 0.997 & 9.6  & 1.006  & 10.9\\  \hline 
{\sc Compass} \cite{ref:compassmult} & $\pi^{\pm}$ (d)& 398 & 1.003 & 378.7 & 1.008 & 382.9  \\                                 

{\sc Hermes} \cite{ref:hermesmult}  & $\pi^{\pm}$ (p) & 64 & 0.981 & 74.0 & 0.986 & 69.9 \\
                               & $\pi^{\pm}$ (d)&   64 &0.980 & 107.3 & 0.985  &103.7\\ \hline 
{\sc Phenix} \cite{ref:phenixdata}   & $\pi^0$ &   15 & 1.174 & 14.3 & 1.167  &14.4  
    \\
{\sc Star} \cite{ref:stardata13} &
                                $\pi^{\pm}$, $\pi^0$  & 38  &1.205   & 31.2 & 1.202 &33.8   \\    
{\sc Alice} \cite{ref:alicedata} \hfill& $\pi^0$    &  11 & 0.696 & 33.3 & 0.700 &31.2     \\ \hline\hline
{\bf TOTAL:} && 924 & & 966.4&  & 875.8 \\
\end{tabular}
\end{ruledtabular}
\end{table}

In Tab.~\ref{tab:exppiontab} we compare the quality of a fit performed within the 
standard ZMVFN factorization scheme, and a GMVFN variant where the prescriptions 
of Eqs.(\ref{massivemix}) and (\ref{masslessmix}) have been adopted for the charm 
and bottom coefficients respectively, as in the example of Fig.\ref{fig:cb_evo}. Among the different
prescriptions we have explored, the above mentioned one reproduces best the data, 
significantly better that in the ZMVFN scheme, with much lower $\chi^2_i$ values 
and smaller normalization shifts $N_i$. No significant improvement is found with 
more sophisticated weight functions $f(Q)$. On the other hand, the most simple 
subtraction of Eq.(\ref{matchingpartoncs}), produces fits of much poor quality,
suggesting that such prescription oversubtracts for charm, and converges much 
slower to the massless limit than the data require for bottom.

\begin{figure}[b!]
\vspace*{-0.8cm}
\hspace*{-6mm}
\epsfig{figure=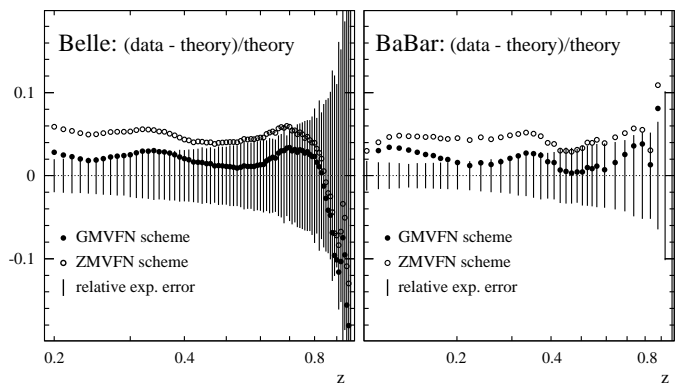,width=0.56\textwidth}
\vspace*{-0.5cm}
\vspace*{-4.5cm}

\caption{Comparison between data from Belle and BaBar and estimates from the
ZMVFN and GMVFN schemes}
\label{fig:belle&babar}
\end{figure}
As expected, heavy quark mass effects are most noticeable for Belle and Babar 
experiments, since at their relatively low center of mass energies heavy quarks
are far from behaving as massless. On the other hand, even though these experiments 
are above the bottom threshold $2m_b$, where the bottom should be considered 
as an active flavor in the ZMVFN scheme, the bottom is actually strongly suppressed, 
feature that is well accounted for in the GMVFN scheme. 
In Fig.\ref{fig:belle&babar} we show the differences between the fit estimates
and Belle and BaBar data, against the relative experimental error.

\begin{figure}[t!]
\vspace*{-0.4cm}
\hspace*{-6mm}
\epsfig{figure=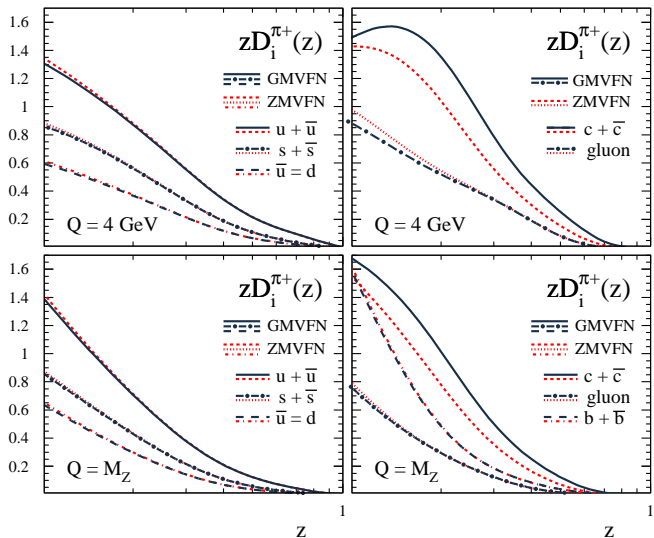,width=0.54\textwidth}
\vspace*{-0.5cm}
\caption{FFs at $Q=4$ GeV and $Q=M_Z$ coming from the ZMVFN and
GMVFN schemes}  
\label{fig:distri_lin}
\end{figure}

Notice that there is also a considerable improvement in the description of data 
from experiments at a higher energy scale. The difference between the massless and
the mass dependent picture comes in this case from the fact that the QCD scale 
dependence preserves the difference between the charm fragmentation probabilities 
constrained by lower energy data, as shown in Fig.\ref{fig:distri_lin}. The bottom
fragmentation probability is mainly constrained by high energy flavor tagged SIA 
data, for which mass dependent corrections become negligible, and therefore the 
results for this flavor in both pictures agree. The GMVFN however guarantees 
that the bottom contribution at lower energies is conveniently suppressed, improving 
the overall agreement and consistency. No significant differences are found for the 
light flavors, that are constrained mainly by light flavor tagged SIA and SIDIS data. 

{\it Conclusions and outlook.---}
We have shown that an accurate determination of the fragmentation probabilities of 
quarks and gluons into pions, matching the precision of the present generation of 
hadroproduction experiments, requires a picture sensitive to heavy quarks dynamics.
Such a picture was presented here, together with the results of a NLO QCD global 
analysis where it was implemented. Heavy quark mass dependence is specially relevant 
in single inclusive electron-positron annihilation into pions, where the detailed 
energy scale dependence of the charm contribution and the suppression of the bottom 
above their respective thresholds is non negligible. These effects are expected to 
be even more apparent in the production of heavy flavored mesons, where the mass 
dependent fragmentation probabilities are dominant. 


We warmly acknowledge Daniel de Florian and Marco Stratmann for help and encouragment.
This work was supported by CONICET, ANPCyT and UBACyT.        


\end{document}